\documentclass[12pt]{article}

\usepackage[english]{babel}
\usepackage[utf8]{inputenc}
\usepackage[toc,page]{appendix}
\usepackage{amsmath,amssymb}
\usepackage{graphicx}
\usepackage{cite}
\usepackage{color}
\usepackage{braket}
\usepackage{subfigure}
\usepackage[colorinlistoftodos]{todonotes}
\numberwithin{equation}{section}
\title{}
\author{}
\date{}
\allowdisplaybreaks

\def\half{{1\over 2}}
\def\del{\partial}

\def\nn{\nonumber}

\def\mM{\mathcal{M}}

\begin{document}

\begin{titlepage}
	\rightline{\vbox{   \phantom{ghost} }}
	
	\begin{flushright}
YITP-19-96
\end{flushright}

	\vskip 1.8 cm
	\begin{center}
		{\Large \bf \boldmath
		Nearly AdS$_2$ holography in quantum CGHS model}
	\end{center}
	\vskip .5cm

	\title{}
	\date{\today}
	\author{Shinji Hirano, Yang Lei, Sam Leuven}

	\centerline{\large {{\bf Shinji Hirano$^{* \dagger}$, Yang Lei$^{*\flat}$}}}
	
	\vskip 1.0cm
	
	\begin{center}
		
		\sl $^*$ 
		School of Physics and Mandelstam Institute for Theoretical Physics\\
		DST-NRF Centre of Excellence in Mathematical and Statistical Sciences (CoE-MaSS)\\
		National Institute for Theoretical Physics\\
		University of the Witwatersrand, WITS 2050, Johannesburg, South Africa, \\
		$^{\dagger}$Center for Gravitational Physics,
		Yukawa Institute for Theoretical Physics\\
		Kyoto University, Kyoto 606-8502, Japan 
		\\
			$^{\flat}$The Niels Bohr Institute, Copenhagen University,\\
			Blegdamsvej 17, DK-2100 Copenhagen \O , Denmark

	\end{center}

	\vskip 1.3cm \centerline{\bf Abstract} \vskip 0.2cm \noindent
	
In light of recent developments in nearly AdS$_2$ holography, we revisit the semi-classical version of two-dimensional dilaton gravity proposed by Callan, Giddings, Harvey, and Strominger (CGHS) \cite{Callan:1992rs} in the early 90's. In distinction to the classical model, the quantum-corrected CGHS model has an AdS$_2$ vacuum with a constant dilaton. By turning on a non-normalizable mode of the Liouville field, i.e. the conformal mode of the 2$d$ gravity,
the explicit breaking of the scale invariance renders the AdS$_2$ vacuum nearly AdS$_2$. As a consequence, there emerges an effective one-dimensional Schwarzian-type theory of pseudo Nambu-Goldstone mode -- the boundary graviton -- on the boundary of the nearly AdS$_2$ space. 
We go beyond the linear order perturbation in non-normalizable fluctuations of the Liouville field and work up to the second order.
As a main result of our analysis, we clarify the role of the boundary graviton in the holographic framework and show the Virasoro/Schwarzian correspondence, namely that the 2$d$ bulk Virasoro constraints are equivalent to the graviton equation of motion of the 1$d$ boundary theory, at least, on the $SL(2,R)$ invariant vacuum.

\end{titlepage}	
\tableofcontents

\section{Introduction}

The AdS$_2$ space makes a universal appearance in the near-horizon limit of extremal black holes. The AdS/CFT correspondence  \cite{Maldacena:1997re, Gubser:1998bc, Witten:1998qj} can be successfully applied to the counting of degeneracy of microstates for extremal black holes \cite{Sen:2008yk,Sen:2011cn}. 
However, from the viewpoint of holography, since there cannot be finite energy excitations in asymptotically AdS$_2$ spaces due to large long-distance backreactions \cite{Maldacena:1998uz}, there is no dynamics in AdS$_2$/CFT$_1$ and what we can learn from it is only degeneracy of ground states. From the viewpoint of black hole physics, it is important to go beyond extremality in order to study black hole evaporations and the information paradox.   

To address these issues, nearly AdS$_2$  (NAdS$_2$) holography was pioneered by Almheiri and Polchinski \cite{Almheiri:2014cka}: The conformal invariance of AdS$_2$ was broken by an introduction of an explicit energy scale and the holographic study of nearly AdS$_2$ geometry was initiated for a class of 2$d$ dilaton gravity models 
\begin{equation}
S_{\rm dilG} = \int d^2 x \sqrt{-g} \left[\Phi R+U(\Phi) (\nabla \Phi)^2-2V(\Phi)\right]
\end{equation}
in which backreactions due to the symmetry breaking scale are under control and can be studied analytically. 
The functions $U$ and $V$ of the dilaton $\Phi$ specify the models of one's interest. 
The Jackiw-Teitelboim (JT) model \cite{Jackiw:1984je,Teitelboim:1983ux}, which has been most studied in recent developments, is given by the choice $U=0$ and $V(\Phi)=\Phi$, whereas the (classical) CGHS model \cite{Callan:1992rs}, whose semi-classical version is of our interest, corresponds to $U(\Phi)= 1/\Phi$ and  $V(\Phi) = -2\lambda^2 \Phi$. As an indication of physics of near-extremal black holes, the JT model, for example, captures the first order correction $\kappa^{-1} T$ to the entropy of near extremal black holes, $S(T) =S_0 +\kappa^{-1} T + \mathcal{O}(T^2)$, where $S_0$ is the entropy of extremal black holes and $\kappa$ is the energy scale of symmetry breaking \cite{Almheiri:2014cka, Maldacena:2016upp, Engelsoy:2016xyb}.

More recent developments have been boosted by the connection between the NAdS$_2$ gravity and the SYK model \cite{Sachdev:2015efa, Maldacena:2016upp, Engelsoy:2016xyb, Maldacena:2016hyu, Kitaev:2017awl,Sarosi:2017ykf}.\footnote{Among many other interesting papers, partial references include \cite{Moitra:2018jqs,Goel:2018ubv,Nayak:2018qej,Cotler:2016fpe, Krishnan:2016bvg,Balasubramanian:2016ids,Jensen:2016pah}. We apologize for our ignorance about any other important works.} 
The latter is an exactly solvable quantum many-body system with an emergent near conformal invariance \cite{Kitaev:KITP,Sachdev:1992fk}. Both are related to black hole physics in higher dimensions. In fact, the  SYK  model saturates the quantum chaos bound which is believed to be a smoking gun for the existence of gravity duals  \cite{Maldacena:2015waa}. Moreover, the boundary effective theory of the NAdS$_2$ gravity has turned out to be a Schwarzian theory which also emerges in the soft sector of the SYK model.


In light of these developments in nearly AdS$_2$ holography, in this paper, we revisit a quantum-corrected version of the CGHS model \cite{Callan:1992rs} as an alternative to the JT model. The classical CGHS model receives a quantum correction due to conformal anomaly described by the well-known non-local Polyakov action \cite{Polyakov:1981rd}. For a large number $N$ of massless scalars, the CGHS model, including the anomaly correction, can be studied semi-classically.  
For our convenience at the risk of being a misnomer, we refer to it as the quantum CGHS (qCGHS) model. 
It has been known that the qCGHS model has an exact AdS$_2$ vacuum with a constant dilaton \cite{Birnir:1992by}. This offers us an opportunity to study the NAdS$_2$ gravity in the qCGHS model.


In the JT model, the scale of symmetry breaking was introduced by the dilaton deformation which grows near the boundary of the AdS$_2$ space and renders the AdS$_2$ vacuum nearly AdS$_2$ \cite{Almheiri:2014cka, Maldacena:2016upp, Engelsoy:2016xyb}. In the qCGHS model, in contrast, the dilaton is a constant and the scale is instead introduced by turning on a non-normalizable mode of the Liouville field, i.e. the conformal mode of the 2$d$ gravity, in much the same way as in the Liouville theory  studied in this context in \cite{Mandal:2017thl}. As a consequence, there emerges a Schwarzian theory on the boundary as in the case of the JT model and the Liouville theory. The Schwarzian theory is an effective theory of pseudo Nambu-Goldstone boson -- the boundary graviton -- associated with the spontaneous breaking of the reparametrization symmetry down to the $SL(2,R)$ subgroup in which the explicit symmetry breaking scale renders the effective action finite in a similar way to the QCD chiral Lagrangian with the pion decay constant.

Owing to the solvability of the Liouville equation,  we are able to study the non-normalizable mode beyond the linear order. We can, in principle, go to arbitrary higher orders, but we content ourselves with working up to the second order in detail. As a main result of our analysis, we clarify the role of the boundary graviton in the holographic framework, which is a degree of freedom somewhat atypical in the standard holography. As we will show, the graviton equation of motion of the 1$d$ boundary theory is equivalent to the 2$d$ bulk Virasoro constraints, at least, on the $SL(2,R)$ invariant vacuum.


This paper is organized as follows: In Section \ref{sec:qCGHS} we will give a brief review on the qCGHS model and its exact AdS$_2$ vacuum as well as more general solutions on which our discussions that follow are based. We will then begin to study nearly AdS$_2$ holography in the qCGHS model in Section 3.
We will first discuss the non-normalizable mode of the Liouville field which renders the AdS$_2$ vacuum nearly AdS$_2$. We will then construct the fully-backreacted NAdS$_2$ geometry and use it to find the 1$d$ boundary effective action up to the second order in the non-normalizable Liouville fluctuation. 
In Section \ref{sec:correspondence}, by using the boundary action derived in Section 3, we will show the (conditional) equivalence between the 2$d$ bulk Virasoro constraints  and  the graviton equation of motion of  the 1$d$ boundary theory. Many of the computational details will be relegated to Appendices \ref{computations} and \ref{computationsII}. Finally,  we will discuss our results and conclude with directions for the future work in Section \ref{sec:discussion}.

\section{The quantum CGHS model}\label{sec:qCGHS}

The CGHS model \cite{Callan:1992rs} is a model of 2$d$ dilaton gravity which arises as the effective two-dimensional theory of extremal dilatonic black holes in four and higher dimensions \cite{Gibbons:1982ih, Gibbons:1987ps, Garfinkle:1990qj, Horowitz:1991cd, Giddings:1991mi, Giddings:1992kn} and is defined by the action
\begin{equation}\label{CGHS}
S_{\rm CGHS} = \frac{1}{2\pi} \int d^2 x \sqrt{-g} \left[e^{-2\phi}(R+ 4(\nabla \phi)^2 +4\lambda^2) -\frac{1}{2} \sum_{i=1}^N (\nabla f_i)^2 \right]\ ,
\end{equation}  
where $g$, $\phi$ and $f_i$ are the metric, dilaton and massless matter fields, respectively, and $\lambda^2$ is a cosmological constant. The matter fields $f_i$ originate from Ramond-Ramond fields in type II superstring theories.

This model has been extensively studied in the early 90's as a model of evaporating black holes. Remarkably, the model is classically solvable and has a simple eternal black hole solution in an asymptotically flat and linear dilaton spacetime. Moreover, it can describe a formation and the subsequent evaporation of the black hole, and it was hoped that significant insights into information paradox might be gained by studying this model and its variants.  
See, for example, for the review \cite{Thorlacius:1994ip, Kazama:1994se}.

Quantum mechanically, the classical action \eqref{CGHS} is corrected by conformal anomaly described by the well-known non-local Polyakov action \cite{Polyakov:1981rd}
\begin{equation}\label{eq:Polyakovaction}
S_P=  -\frac{N-24}{96 \pi} \int d^2x \int d^2 y \sqrt{-g(x)} \sqrt{-g(y)} R(x)\frac{1}{\square} R(y)
\end{equation}
where $N-24=(N+2)-26$: $N$ is due to the massless matter fields, $2$ from the dilaton and the conformal mode of the 2$d$ metric, and $-26$ from the diffeomorphism $bc$ ghosts.\footnote{We will eventually focus on the large $N$ limit in which we can ignore $-24$.} 
Thus the quantum-corrected CGHS model is defined by the action 
\begin{equation}\label{eq:qCGHSaction}
S_{\text{qCGHS}} = S_{\rm CGHS}+S_P\ .
\end{equation}
To be precise, this is a semi-classical version of the CGHS model. Nevertheless, for our convenience, we shall refer to it as quantum CGHS model (qCGHS) in the rest of the paper.

In the conformal gauge 
\begin{equation}\label{confgauge}
ds^2 =-e^{2\rho } dx^{+} dx^-\qquad\mbox{where}\qquad x^{\pm}=x^0\pm x^1\ ,
\end{equation}
the non-local Polyakov action becomes local and is given by the Liouville action, and the qCGHS action takes the form
\begin{align}\label{eq:qCGHSactiongauge}
\hspace{-.2cm}
S_{\rm qCGHS}= \int {d^2x\over\pi} \left[ e^{-2\phi} \left(2\partial_+\partial_- \rho -4 \partial_+\phi \partial_-\phi +\lambda^2 e^{2\rho} \right)
-\frac{N}{12} \partial_+\rho \partial_- \rho + \frac{1}{2}\sum_{i=1}^N \partial_+f_i \partial_- f_i \right].
\end{align} 
Here and hereafter we consider the large $N$ limit in which $N-24$ can be replaced by $N$. 
The equations of motion for the Liouville field $\rho$, dilaton $\phi$ and matter fields $f_i$ are given, respectively, by
\begin{align}
0 &= T_{+-} = e^{-2\phi} (2\partial_+\partial_-\phi -4\partial_+\phi \partial_-\phi -\lambda^2 e^{2\rho}) -\frac{N}{12} \partial_+\partial_- \rho\ , \label{qCGHSEOMliouville}\\ 
0&= -4\partial_+\partial_- \phi +4\partial_+\phi \partial_- \phi +2\partial_+\partial_-\rho +\lambda^2 e^{2\rho}\ ,\label{qCGHSEOMdilaton}\\
0&=\partial_+\partial_- f_i\ .\label{qCGHSEOMmatter}
\end{align}
In addition, this system is subjected to the Virasoro constraints, i.e. the equations of motion for $g_{\pm\pm}$: 
\begin{align}\label{Virasoro}
0 =T_{\pm\pm} = e^{-2\phi} (4\partial_\pm\phi \partial_\pm \rho -2\partial^2_\pm \phi) +\frac{1}{2} \sum_{i=1}^N\partial_\pm f_i \partial_\pm f_i -\frac{N}{12} (\partial_\pm \rho \partial_\pm \rho -\partial_\pm^2 \rho +t_\pm)\ .
\end{align}
The last quantities $t_\pm$ reflect the non-locality of the Polyakov action and are determined by the choice of the vacuum.\footnote{An elegant and convenient way to see it explicitly is to introduce an auxiliary field $\varphi$ obeying $\square \varphi =R$ in terms of which the non-local Polyakov action can be rewritten as $S_P= \frac{N}{96 \pi} \int d^2x \sqrt{-g} (-\varphi \square \varphi +2\varphi R)$ \cite{Hayward:1994dw,fabbri2005modeling}. 
In the conformal gauge \eqref{confgauge} the equation of motion yields
$\varphi=-2\rho + 2\varphi_+(x^+)+2\varphi_-(x^-)$
with arbitrary (anti-)holomorphic functions $\varphi_{\pm}(x^{\pm})$. The energy-moment tensor in this reformulation is given by the last Liouville part of   \eqref{Virasoro} with  
$t_\pm =\partial_\pm^2 \varphi_\pm -(\partial_\pm \varphi_\pm)^2$. }\label{foot:auxiliary}

The quantum CGHS model is no longer solvable and there is no simple analytic black hole solution even though there is a modified solvable variant of the qCGHS model known as the RST model proposed in \cite{Russo:1992ax, Thorlacius:1994ip} and extensively studied thereafter. 
For the purpose of holography, however, we are interested in asymptotically AdS$_2$ spacetimes. Indeed, there exists an AdS$_2$ vacuum with a constant dilaton in the quatum CGHS model \cite{Birnir:1992by}: 
\begin{equation}
\rho =  \ln \left[ \frac{\sqrt{2}}{\lambda(x^+- x^-)}\right]\ , \qquad e^{2\phi}= \frac{24}{N}\ , \qquad\mbox{and}\qquad t_\pm =0\ ,
\end{equation}
where $x^\pm= t\pm z$ are the lightcone coordinates in the Poincar\'e patch of AdS$_2$. 

Moreover, there exist a more general class of solutions obtained by the reparametrizations 
$x^+\mapsto A(x^+)$ and $x^-\mapsto B(x^-)$ \cite{Seiberg:1990eb}:
\begin{align}
\rho &= \frac{1}{2}\ln \frac{2A'(x^+)B'(x^-)}{\lambda^2(A(x^+)-B(x^-))^2}\ , \qquad\quad e^{2\phi} = \frac{24}{N}\ ,\label{eq:generalABsolution}\\
t_+ &={1\over 2}\{A(x^+), x^+\}\ , \qquad\quad t_- = {1\over 2}\{B(x^-), x^-\}\ ,\label{tplusminus}
\end{align}
where we introduced the Schwarzian derivative defined by
\begin{align}
\{f(\zeta), \zeta\}:= \frac{f'(\zeta)f'''(\zeta)-{3\over 2}f''(\zeta)^2}{f'(\zeta)^2}\ .
\end{align}
Note that the choice of $t_{\pm}$ corresponds to $\varphi_+(x^+)={1\over 2}\ln A'(x^+)$ and $\varphi_-(x^-)={1\over 2}\ln B'(x^-)$ in footnote \ref{foot:auxiliary}.

\section{Nearly AdS$_2$ holography in qCGHS model}

The AdS$_2$ space appears universally in the near horizon limit of extremal black holes as a 2-dimensional component of higher dimensional spacetimes. 
In contrast to higher dimensional counterparts, however, the AdS$_2$ boundary conditions are not consistent with finite energy excitations due to large long-distance backreactions \cite{Maldacena:1998uz}. From the black hole viewpoint, a mass gap is developed in the near horizon region and the AdS$_2$/CFT$_1$ correspondence can only describe the ground state degeneracy. In order to have nontrivial dynamics, one must therefore introduce a new scale and enforces a deviation from the pure AdS$_2$ space which does not die off near the boundaries. This necessitates turning on a non-normalizable mode dual to an irrelevant operator in conformal mechanics. From the extremal black hole perspective, this deformation effectively undoes the near horizon decoupling and enables excursions into the region of spacetime corresponding to UV of the dual field theory.    

To realize this scenario, we first cut off the AdS$_2$ space near its boundary at a small finite $z$. More precisely, we consider the spacetime \eqref{eq:generalABsolution} with $A=B$.
The resulting spacetime is a reparametrization of the Poincar\'e AdS$_2$ by
\begin{align}
(t, z)\quad\mapsto\quad (\tilde{t}, \tilde{z})\equiv \left({1\over 2}(B(t+z)+B(t-z)), {1\over 2}(B(t+z)-B(t-z))\right)
\end{align}
which near the boundary becomes
\begin{align}
(\tilde{t}, \tilde{z})=\left(B(t), B'(t)z\right)+\mathcal{O}(z^2)\ ,
\end{align}
where the map $t\mapsto B(t)$ is the time reparametrization on the cutoff boundary.
It is, however, important to note that $B(t)$ is not a mere time reparametrization but physical: 
A different reparametrization function $B(t)$ results in a different $t_\pm$ in \eqref{Virasoro} and \eqref{tplusminus}. In other words, a change to $B(t)$ results in a change to the vacuum or the boundary condition. This then implies that physical observables such as correlation functions do depend on $B(t)$. 
Note, however, that there is a subset of $B(t)$ for which $t_\pm =0$:
\begin{equation}\label{eq:sl2stran}
B(t) = \frac{ a t+b}{ct +d}\qquad\quad\mbox{with}\qquad\quad ad-bc = 1\ .
\end{equation}
This is a M\"obius transformation of $t$. It can be interpreted as meaning that the reparametrization symmetry is spontaneously broken to $SL(2,R)$ and $B(t)$ is the Nambu-Goldstone boson associated with the broken symmetry.  

In the meantime, the conformal factor of the metric \eqref{eq:generalABsolution} has the boundary expansion 
\begin{equation}\label{eq:asymAdS2}
2\lambda^2e^{2\rho} = \frac{1}{z^2} + {1\over 6}\{B(t), t\} +\mathcal{O}(z^2)\ .
\end{equation}
The finite part in the expansion reminds us of the Brown-Henneaux asymptotics of the AdS$_3$ space \cite{Brown:1986nw} and it may thus provide another perspective: $B(t)$ can be thought of as the boundary graviton living in the cutoff surface at a small $z$ \cite{Maldacena:2016upp}. 

\subsection{Non-normalizable mode and symmetry breaking scale}\label{sec:NN}

In the case of Jackiw-Teitelboim (JT) gravity \cite{Jackiw:1984je, Teitelboim:1983ux}, the new scale $\phi_r$ to deform the AdS$_2$ vacuum is introduced through the dilaton which grows as $\phi\sim\phi_r/z$ near the boundary \cite{Maldacena:2016upp}.
In contrast, as we will see,  in the case of the qCGHS model, the dilaton plays only a minor role and the new scale, which renders the AdS$_2$ vacuum nearly AdS$_2$, is provided by a non-normalizable mode of the Liouville field $\rho$. This is very much similar to the mechanism advocated in \cite{Mandal:2017thl}. (A related idea was discussed in an earlier literature \cite{Nakayama:2017pys}.) 

Whether it is the dilaton $\phi\equiv\phi_0+\tilde{\phi}$ or the Liouville field $\rho\equiv\rho_0+\tilde{\rho}$, since what is essential for the nearly AdS$_2$ geometry is the non-normalizable mode, we first analyze the fluctuations $\tilde{\phi}$ and $\tilde{\rho}$ of the dilaton and Liouville fields in the qCGHS model. For this purpose, we work in the conformal gauge \eqref{confgauge} and 
then the quadratic fluctuation action for the dilaton-Liouville system is given by
\begin{align}
\hspace{-0.4cm}
S^{(2)}_{\phi+\rho}=&{N\over 12\pi}\int_{\mM}dx^+dx^- \biggl[-\del_+\tilde{\phi}\del_-\tilde{\phi}-\lambda^2 e^{2\rho_0}\tilde{\phi}^2-\del_+(\tilde{\rho}-\tilde{\phi})\del_-(\tilde{\rho}-\tilde{\phi})
+\lambda^2 e^{2\rho_0}(\tilde{\rho}-\tilde{\phi})^2\biggr],
\end{align}
where as in \eqref{eq:generalABsolution} the background Liouville and dilaton fields are
\begin{align}
\rho_0 &= \frac{1}{2}\ln \frac{2B'(x^+)B'(x^-)}{\lambda^2(B(x^+)-B(x^-))^2}\ , \qquad\quad e^{2\phi_0} = \frac{24}{N}\ .
\end{align}
The fluctuation fields are thus classified into the ``tachyonic'' dilaton $\tilde{\phi}$ and the massive field $\tilde{\rho}-\tilde{\phi}$ besides $N$ massless matter fields $f_i$.\footnote{In the previous version of our paper, we missed the mass term for the dilaton $\tilde{\phi}$  and stated wrongly that it is massless.} 
It needs to be mentioned that the dilaton fluctuation violates the Breitenlohner-Freedman bound \cite{Breitenlohner:1982bm, Breitenlohner:1982jf}. Noting, however, that it behaves as $\tilde{\phi}\sim \sqrt{z}\cos(\sqrt{7}/2\log z)$ for all real and imaginary frequencies near the boundary $z=0$ of the Poincar\'e AdS$_2$, the linear instability can be alleviated by imposing the Neumann boundary condition $\del_z\phi(t, z)=0$ at the boundary that freezes the dilaton fluctuation.

Having frozen the dilaton fluctuation by the Neumann boundary condition, we now focus on the massive Liouville fluctuation $\tilde{\rho}$.\footnote{The massless matter fields $f_i$ are dual to marginal operators of conformal dimension $\Delta=1$, whereas, as we will see, the massive field $\tilde{\rho}-\tilde{\phi}$ is dual to an irrelevant operator of dimension $\Delta=2$.}
To illustrate the essential point, we first consider  the Poincar\'e AdS$_2$ corresponding to $B(x^{\pm})=t\pm z$. 
The  equation of motion for the Liouville fluctuation is then
\begin{equation}\label{LiouflucEqPoincare}
\left[ -\partial_t^2+\partial_z^2 -\frac{2}{z^2}\right] \tilde{\rho}=0\ .
\end{equation} 
Near the boundary the solution to this equation goes as $\tilde{\rho}\sim \alpha/z+\beta z^2$, which indicates that the Liouville fluctuation $\tilde{\rho}$ is dual to an irrelevant operator of conformal dimensions $\Delta = 2$. To be more precise, the non-normalizable mode is given by \cite{Witten:1998qj}
\begin{equation}
\tilde{\rho}(t,z) = - \int_{-\infty}^{\infty} dt_0 \frac{2z^2j_{\rho}(t_0)}{\left(z^2-(t-t_0)^2\right)^2} \ ,
\end{equation}
where a particular normalization was chosen for the consistency with the analysis that follows.
We would like to emphasize that the source $j_{\rho}$ is an analogue of $\phi_r$ in JT gravity and the advertised new length scale which renders the AdS$_2$ vacuum nearly AdS$_2$. We thus anticipate that the finite  action for the pseudo Nambu-Goldstone boson $B(t)$ is schematically of the form  \cite{Maldacena:2016upp, Mandal:2017thl}
\begin{align}\label{pNG}
S_{\rm pNG}\sim\int dt j_{\rho}(t)\{B(t), t\}+\cdots
\end{align}
where the source $j_{\rho}$ is the explicit symmetry breaking scale and an analogue of the pion decay constant. We will make it more precise in Section \ref{Sec:Action}.

For a generic $B$, the fluctuation equation \eqref{LiouflucEqPoincare} is generalized to
\begin{align}\label{1stflucgen}
\left[\del_+\del_-  +\frac{2B'(x^+)B'(x^-)}{\left(B(x^+)-B(x^-)\right)^2}\right] \tilde{\rho}=0
\end{align}
and the non-normalizable mode is 
\begin{align}\label{NNLiouville}
\tilde{\rho}(x^+,x^-)=-\half\int_{-\infty}^{\infty}dB(t_0)\left({1\over B(x^+)-B(t_0)}-{1\over B(x^-)-B(t_0)}\right)^2\tilde{j}_{\rho}(B(t_0))\ .
\end{align}
Note that under the reparametrization $t\mapsto B(t)$, since the source $j_{\rho}$ transforms according to $j_{\rho}(t)dt^{-\Delta+1}=\tilde{j}_{\rho}(B(t))(B'(t)dt)^{-\Delta+1}$, the transformed source is related to the one in the Poincar\'e AdS$_2$ by $\tilde{j}_{\rho}(B(t))=j_{\rho}(t)B'(t)$.

Similarly, by solving \eqref{qCGHSEOMmatter}, it is straightforward to find the non-normalizable matter fields
\begin{align}\label{NNmatter}
f_i(x^+, x^-)=\int_{-\infty}^{\infty}dB(t_0)\left({1\over B(x^+)-B(t_0)}-{1\over B(x^-)-B(t_0)}\right)\tilde{j}_{f_i}(B(t_0))\ ,
\end{align}
where the transformed source $\tilde{j}_{f_i}(B(t))$ is related to the one in the Poincar\'e AdS$_2$ by $\tilde{j}_{f_i}(B(t))=j_{f_i}(t)$. Note that as mentioned above, the sources for massless fields do not introduce a scale since it is dual to marginal operators.

\subsection{Nearly AdS$_2$ geometry in qCGHS model}\label{fullbackreaction}

In the previous section we deformed the AdS$_2$ vacuum to the linear order in the non-normalizable Liouville fluctuations. In fact, owing to the solvability of the Liouville equation, one can go beyond perturbation and resum the nearly AdS$_2$ deformation to all orders.
To see it, recall the equations of motion \eqref{qCGHSEOMliouville} and \eqref{qCGHSEOMdilaton}. For a constant dilaton, the two equations reduce to a single equation
\begin{align}\label{SingleLiouvilleEq}
0=2\del_+\del_-\rho+\lambda^2 e^{2\rho}\ .
\end{align}
The general solution is the Liouville field $\rho$ in \eqref{eq:generalABsolution}. The Liouville fluctuation equation \eqref{1stflucgen} is an expansion of this equation to the linear order:
\begin{align}
0=\left[2\del_+\del_-\rho_0+\lambda^2 e^{2\rho_0}\right]+2\left[\del_+\del_-+\lambda^2e^{2\rho_0}\right]\tilde{\rho}+\mathcal{O}(\tilde{\rho}^2)\ .
\end{align}
The non-normalizable mode \eqref{NNLiouville}  thus resums to
\begin{align}\label{resum}
e^{2(\rho_0+\tilde{\rho})}={2\del_+(B(x^+)+b_+(x^+))\del_-(B(x^-)+b_-(x^-))\over\lambda^2\left[(B(x^+)+b_+(x^+))-(B(x^-)+b_-(x^-))\right]^2}
\end{align}
with the deformation 
\begin{align}\label{deformab}
b_{\pm}(x^{\pm})=b(x^{\pm})\equiv\int_{-\infty}^{\infty}dt_0{\tilde{j}_{\rho}(B(t_0))\over B(x^{\pm})-B(t_0)}
\end{align} 
which can be inferred from the expansion of the resummed expression.
This gives the fully-backreacted nearly AdS$_2$ geometry described by the metric $ds_{{\rm NAdS}_2}^2=-e^{2(\rho_0+\tilde{\rho})}dx^+dx^-$.

In order to gain better ideas of this geometry, we consider the nearly Poincar\'e AdS$_2$ corresponding to $B(x^{\pm})=t\pm z$. 
After performing a Wick-rotation, $t\to i\tau$ and $j_{\rho}(t)\to -i\,j_{\rho}(\tau)$, the deformation near the boundary takes a simple form
\begin{align}
b(x^{\pm})=\int_{-\infty}^{\infty}d\tau_0{j_{\rho}(\tau_0)\over i(\tau-\tau_0)\pm z}
\stackrel{z\to 0}{\longrightarrow}\pm\pi j_{\rho}(\tau)-i\int_{-\infty}^{\infty}d\tau_0P\left[{1\over\tau-\tau_0}\right]j_{\rho}(\tau_0)\ .
\end{align}
This amounts to the coordinate transformation
\begin{align}
\tau\mapsto \tau-\int_{-\infty}^{\infty}d\tau_0P\left[{1\over\tau-\tau_0}\right]j_{\rho}(\tau_0)\ ,\qquad\quad
z\mapsto z+\pi j_{\rho}(\tau)\ .
\end{align}
Rather than viewing this as a mere coordinate transformation, 
we may interpret it as meaning that the non-normalizable deformation cuts out the near-boundary region below the symmetry breaking scale $z_{\star}=\pi j_{\rho}(\tau)$ even though the space is locally AdS$_2$.

It should be noted that we have not imposed the Virasoro constraints \eqref{Virasoro}. As we will see in Section \ref{sec:correspondence}, the Virasoro constraints impose a restriction on the functional form of the source $j_{\rho}(t)$. 

\subsection{Second order perturbation}\label{sec:secondorder}

Our next goal is to construct the 1$d$ boundary effective theory of the pseudo Nambu-Goldstone boson $B(t)$ as alluded in \eqref{pNG}. We are going beyond the linear order in $j_{\rho}$ as typically done in the literature and work out to the second order in order to perform a nontrivial check of nearly AdS$_2$ holography in the qCGHS model in Section \ref{sec:correspondence}.

The resummation \eqref{resum} of the non-normalizable mode allows us to systematically extract the Liouville fluctuations higher orders in $j_{\rho}$. 
For the clarity of the argument, we expand the Liouville fluctuation $\tilde{\rho}$ as 
\begin{align}
\tilde{\rho}=\rho_1+\rho_2+\cdots
\end{align} 
where the numbers in the subscript denote the order in the source $j_{\rho}$. In this notation, the first order non-normalizable mode \eqref{NNLiouville} is renamed to 
\begin{align}\label{rho1general}
\rho_1=\half\left[ \frac{b'(x_+)}{B'(x_+)} +  \frac{b'(x_-)}{B'(x_-)}  -\frac{2(b(x_+)-b(x_-))}{B(x_+)-B(x_-)}\right]\ ,
\end{align}
where the RHS of  \eqref{NNLiouville} was rewritten in terms of the deformation \eqref{deformab}.
By expanding \eqref{resum} for a small deformation, one can similarly find the second order Liouville fluctuation
\begin{align}\label{rho2general}
\rho_2=-{1\over 4} \Biggl[\frac{b'(x_+)^2}{B'(x_+)^2}+\frac{b'(x_-)^2}{B'(x_-)^2} -2 \left(\frac{b(x_+)-b(x_-)}{B(x_+)-B(x_-)}\right)^2\Biggr] .
\end{align}
For our purposes, we are interested in the expressions for $\rho_1$ and $\rho_2$ near the boundary at a small $z$. Our strategy is to first find the expressions in the Poincar\'e AdS$_2$ with $B(x^{\pm})=t\pm z$ and then covariantize the results so obtained to reinstate the dependence on $B(t)$. 
We perform an appropriate Wick-rotation, $t\to i\tau$ and $j_{\rho}(t)\to -i\,j_{\rho}(\tau)$, and work in the Euclidean space. The details of the computation are shown in Appendix \ref{computations}.

In the Poincar\'e coordinates, the first order fluctuation is calculated as
\begin{align}
\rho_1=-\pi\left[{j_{\rho}(\tau)\over z}+\half z j_{\rho}''(\tau)\right]+O(z^2)\ .
\end{align}
As discussed in Section \ref{sec:NN}, the divergent term is essential for the appearance of the finite Schwarzian action \eqref{pNG}.
In the meantime, since we work through to the second order in $j_{\rho}$, we would also need the bilinear quantities of $\rho_1$: 
\begin{align}
\hspace{-.05cm}
\rho_1^2=2\pi z\int_{-\infty}^{\infty}  d\tau_0{j_{\rho}(\tau)j_{\rho}(\tau_0)\over (\tau-\tau_0)^4}+\mathcal{O}(z^2)\quad\mbox{and}\quad
\rho_1\del_z\rho_1=\pi \int_{-\infty}^{\infty}  d\tau_0{j_{\rho}(\tau)j_{\rho}(\tau_0)\over (\tau-\tau_0)^4}+\mathcal{O}(z).
\end{align}
To covariantize these expressions, we make the replacements
\begin{align}
z\to zB'(\tau)\ ,\quad \tau\to B(\tau)\ ,\quad\mbox{and}\quad j_{\rho}(\tau)\to \tilde{j}_{\rho}(B(\tau))=j_{\rho}(\tau)B'(\tau)\ .
\end{align}
We thus obtain to the relevant order in $z$ 
\begin{align}\label{rho1boundary}
\rho_1=-\pi{j_{\rho}(\tau)\over z}\ ,\qquad
\rho_1\del_z\rho_1=\pi \int_{-\infty}^{\infty}  dB(\tau_0){B'(\tau)^2B'(\tau_0)j_{\rho}(\tau)j_{\rho}(\tau_0)\over (B(\tau)-B(\tau_0))^4}\ ,
\end{align}
and 
\begin{align}\label{rho1squaredboundary}
{\rho_1^2\over z}=2\pi \int_{-\infty}^{\infty}  dB(\tau_0){B'(\tau)^2B'(\tau_0)j_{\rho}(\tau)j_{\rho}(\tau_0)\over (B(\tau)-B(\tau_0))^4}\ .
\end{align}
These three quantities form a part of the building blocks for the construction of the 1$d$ boundary Schwarzian-type theory.

Turning to the second order fluctuation $\rho_2$, it is similarly calculated as 
\begin{align}
\hspace{-.0cm}
\rho_2= {\pi\over 2z}\int_{-\infty}^{\infty}  d\tau_0{j_{\rho}(\tau)j_{\rho}(\tau_0)\over (\tau-\tau_0)^2}
+\mathcal{O}(z).
\end{align}
The covariant form of the second order fluctuation to the relevant order in $z$ is then found to be
\begin{align}\label{rho2boundary}
\rho_2= {\pi\over 2z}\int_{-\infty}^{\infty}  dB(\tau_0){B'(\tau_0)j_{\rho}(\tau)j_{\rho}(\tau_0)\over (B(\tau)-B(\tau_0))^2}\ .
\end{align}
Apart from $N$ matter fields $f_i$, together with the above three quantities made of $\rho_1$, this forms a complete set of the building blocks for the boundary action we discuss in the next section. 

\subsection{The boundary Schwarzian-type action}\label{Sec:Action}

We are now in a position to discuss the 1$d$ boundary effective theory of the pseudo Nambu-Goldstone boson $B(t)$. We find it most convenient to work in the locally AdS$_2$ gauge adopted in  \cite{Mandal:2017thl}, i.e. factorizing the metric into the background and fluctuation parts:
\begin{align}
ds^2=-e^{2\tilde{\rho}}\left[e^{2\rho_0}dx^+dx^-\right]\equiv e^{2\tilde{\rho}} \hat{g}_{mn}dx^{m}dx^{n}\ .
\end{align}
In this gauge the Liouville $\tilde{\rho}$-dependent part of the non-local Polyakov action becomes \footnote{We set $2\lambda^2 =1$ for simplicity.}
\begin{align}
S_P&={N\over 24\pi}\int_{\cal M} d^2x\sqrt{-\hat{g}}\left[\hat{g}^{mn}\del_{m}\tilde{\rho}\del_{n}\tilde{\rho}+\hat{R}\tilde{\rho}\right]
-{N\over 12\pi}\int_{\del{\cal M}}dt\sqrt{-\hat{\gamma}}\hat{K}\tilde{\rho}\nn\\
&={N\over 48\pi}\int_{\cal M} d^2x\left[-4\del_{+}\tilde{\rho}\del_{-}\tilde{\rho}-2e^{2\rho_0}\tilde{\rho}\right]
+{N\over 12\pi}\int_{\del{\cal M}}dt \tilde{\rho}\del_z\rho_0\ ,
\end{align}
where we used $\hat{R}=-2$ and $\hat{K}=-e^{-\rho_0}\del_z\rho_0$. The last term is a Gibbons-Hawking-York term \cite{Gibbons:1976ue, York:1972sj} for the Liouville theory. 
Now, recall the Liouville equation of motion \eqref{SingleLiouvilleEq}. Its fluctuation part is given by
\begin{equation}
0=2\del_+\del_-\tilde{\rho}+e^{2\rho_0}\left(\tilde{\rho}+\tilde{\rho}^2+\cdots\right)\ .
\end{equation}
From this equation, we can infer to the second order that
\begin{align}
\tilde{\rho}\del_+\del_-\tilde{\rho}=-{1\over 2}e^{2\rho_0}\tilde{\rho}^2+\mathcal{O}(\tilde{\rho}^3)\ ,\qquad\qquad
e^{2\rho_0}\tilde{\rho}=-2\del_+\del_-\tilde{\rho}+2\tilde{\rho}\del_+\del_-\tilde{\rho}+\mathcal{O}(\tilde{\rho}^3)\ .
\end{align}
With the latter on-shell equation and by integration by parts, the Polyakov action simplifies and is only left with the boundary contribution
\begin{align}\label{Polyakovrhotilde}
S_P={N\over 24\pi}\int_{\del{\cal M}} dt\biggl[\tilde{\rho}\del_z\tilde{\rho}-\del_z\tilde{\rho} +2\tilde{\rho}\del_z\rho_0\bigg]\ .
\end{align}
Meanwhile, the classical CGHS action, the first three terms of \eqref{eq:qCGHSactiongauge} in parenthesis, vanishes on-shell and there is only a boundary contribution from the Gibbons-Hawking-York term of the dilaton gravity:
\begin{align}\label{CGHScontribution}
S_{\rm CGHS}+S_{\rm GHY}=-{1\over\pi}\int_{\del{\cal M}} dt\sqrt{-\gamma}e^{-2\phi}K={N\over 24\pi}\int_{\del{\cal M}} dt\del_z(\rho_0+\tilde{\rho})\ ,
\end{align}
where $\gamma=-e^{\rho}$ and $K=-e^{-\rho}\del_z\rho$ and the metric without a hat is the full metric including both the background and fluctuations. 
It is worth noting that this is a contribution genuinely from the qCGHS model. Without this contribution, our analysis, apart from the second order corrections, would virtually have no difference from that of the Liouville theory \cite{Mandal:2017thl}.
Even though the dilaton has been playing only a minor role and this boundary contribution might look rather insignificant, as we will see, it makes important difference in the working precision of nearly AdS$_2$ holography.

At this point we are finding that
\begin{align}
S_{\rm qCGHS}+S_{\rm GHY}={N\over 24\pi}\int_{\del{\cal M}} dt\biggl[\rho_1\del_z\rho_1+\del_z\rho_0 +2(\rho_1+\rho_2)\del_z\rho_0\bigg]
+\mathcal{O}(\tilde{\rho}^3)\ ,
\end{align}
where the background $\rho_0$-dependent term is
\begin{align}\label{delzrho0}
\del_z\rho_0=-{1\over z}+z{2\{B(t),t\}\over 3}\qquad
\stackrel{{\rm Wick-rotation}}{\xrightarrow{\hspace*{1.5cm}}}\qquad
-{1\over z}-z{2\{B(\tau),\tau\}\over 3}\ .
\end{align}
There are $1/z^2$ and $1/z$ divergences in the boundary action we have obtained so far since $\rho_1$ and $\rho_2$ are singular as $1/z$ as discussed in the previous sections. The $1/z^2$ divergences can be removed by adding the boundary cosmological constant as a counter-term following the holographic renormalization procedure \cite{deHaro:2000vlm}:
\begin{align}
S_{ct}={N\over 12\pi}\int_{\del{\cal M}} dt \sqrt{-\gamma}={N\over 12\pi}\int_{\del{\cal M}} dt e^{\rho_0}\left(1+\rho_1+\rho_2+\half\rho_1^2+\cdots\right)\ ,
\end{align}
where the background boundary cosmological constant is
\begin{align}\label{erho0}
e^{\rho_0}={1\over z}+z{\{B(t),t\}\over 3}\qquad
\stackrel{{\rm Wick-rotation}}{\xrightarrow{\hspace*{1.5cm}}}\qquad
{1\over z}-z{\{B(\tau),\tau\}\over 3}\ .
\end{align}
However, there still remains a $1/z$ divergence in $\del_z\rho_0+2e^{\rho_0}$. As it turns out, this is cancelled by the background part of the non-local Polyakov action which we have omitted so far:
\begin{align}
\bar{S}_P&=-{N\over 96\pi}\left[\int_{\cal M} d^2x\sqrt{-\hat{g}}\hat{R}{1\over\widehat{\Box}}\hat{R}-2\int_{\del{\cal M}} dt\sqrt{-\hat{\gamma}}\hat{K}{1\over\widehat{\Box}}\hat{R}\right]\nn\\
&=-{N\over 24\pi}\left[\int_{\cal M} dtdz e^{2\rho_0}\rho_0-\int_{\del{\cal M}} dt(\del_z\rho_0)\rho_0\right]
= -{N\over 24\pi}\int_{\del{\cal M}} dt{1\over z}+\mathcal{O}(z)\ .
\end{align}
We now put all the pieces together to obtain the finite boundary action 
\begin{align}
S_{\rm pNG}={N\over 24\pi}\int_{\del{\cal M}} dt\biggl[\rho_1\del_z\rho_1+e^{\rho_0}\rho_1^2 +2(\rho_1+\rho_2)\left(\del_z\rho_0+e^{\rho_0}\right)\bigg]
+\mathcal{O}(\tilde{\rho}^3)\ .
\end{align}
Note that the second term is a finite contribution that comes from the counter-term $S_{ct}$ and corresponds to a double trace deformation considered in \cite{Klebanov:1999tb}.

Using the expressions \eqref{rho1boundary}, \eqref{rho1squaredboundary} and \eqref{rho2boundary} for the fluctuations together with the background values \eqref{delzrho0} and \eqref{erho0}, after the Wick-rotation, the second order boundary action for the pseudo Nambu-Goldstone boson becomes
\begin{align}\label{pNGfinal}
\boxed{S_{\rm pNG}={N\over 24}\left(2S_{j_{\rho}{\rm Sch}}+3S_{j_{\rho}^2}-S_{j_{\rho}^2{\rm Sch}}\right)}
\end{align}
where we defined 
\begin{align}
S_{j_{\rho}{\rm Sch}}&=\int_{-\infty}^{\infty}d\tau j_{\rho}(\tau)\{B(\tau), \tau\}\ ,\label{Schwarzianaction}\\
S_{j_{\rho}^2}&=\int_{-\infty}^{\infty}d\tau\int_{-\infty}^{\infty}d\tau_0{B'(\tau)^{2}B'(\tau_0)^{2}j_{\rho}(\tau)j_{\rho}(\tau_0)\over (B(\tau)-B(\tau_0))^{4}}\ ,
\label{rhotwopoint}\\
S_{j_{\rho}^2{\rm Sch}}&=\int_{-\infty}^{\infty} d\tau \int_{-\infty}^{\infty}d\tau_0{B'(\tau_0)^2j_{\rho}(\tau)j_{\rho}(\tau_0)\over \left(B(\tau)-B(\tau_0)\right)^2}\{B(\tau), \tau\}
\ .\label{rho2Schaction}
\end{align}
The first action $S_{j_{\rho}{\rm Sch}}$ is the Schwarzian action found in \cite{Maldacena:2016upp, Engelsoy:2016xyb, Mandal:2017thl} as expected. The second action $S_{j_{\rho}^2}$ comes from the quadratic terms in $\rho_1$ and, in the standard holography, corresponds to the two-point function of a dimension $\Delta=2$ operator. Meanwhile, the third action  $S_{j_{\rho}^2{\rm Sch}}$ is the one from the second order fluctuation $\rho_2$ and is a reflection of the fact that the $\Delta=2$ Schwarzian operator, dual to the Liouville fluctuation $\tilde{\rho}$, is a quasi-primary rather than primary.

To be complete, we shall add the massless matter action. By integration by parts and using the equation of motion \eqref{qCGHSEOMmatter}, the matter action becomes
\begin{align}
S_f={1\over 4\pi}\sum_{i=1}^N\int_{-\infty}^{\infty} d\tau f_i\del_zf_i\ .
\end{align}
Thus the boundary action for the non-normalizable mode \eqref{NNmatter} is found to be
\begin{align}\label{matteraction}
S_f={1\over 2}\sum_{i=1}^N\int_{-\infty}^{\infty}d\tau\int_{-\infty}^{\infty}d\tau_0{B'(\tau)B'(\tau_0)j_{f_i}(\tau)j_{f_i}(\tau_0)\over (B(\tau)-B(\tau_0))^{2}}\ ,
\end{align}
where the details of the computation are shown in Appendix \ref{computations}. This is of the form of the two-point function of dimension $\Delta=1$ operators as expected.

As a final note in this section, in the case of  nearly AdS$_2$ holography \cite{Almheiri:2014cka, Maldacena:2016upp, Engelsoy:2016xyb}, it is rather remarkable that the 1$d$ boundary theory is directly ``derived'' from the 2$d$ bulk gravity in the sense that the boundary effective action \eqref{pNGfinal} plus \eqref{matteraction} is expected to be a collective field description of a 1$d$ quantum mechanical theory such as the SYK model \cite{Kitaev:KITP, Sachdev:1992fk}. 

\section{The Virasoro/Schwarzian correspondence}\label{sec:correspondence}

In the standard holography, the 1$d$ boundary effective action
\begin{align}\label{effectiveaction}
S^{({\rm 1d})}_{\rm eff}[B, j]\equiv S_{\rm pNG}+S_f
\end{align}
is interpreted as the generating functional of correlation functions of the operators dual to the sources $j_{\rho}$ and $j_{f_i}$ \cite{Gubser:1998bc, Witten:1998qj}.
However, this is not the end of the story for nearly AdS$_2$ holography: 
As remarked in the end of Section \ref{fullbackreaction}, we have not imposed the Virasoro constraints \eqref{Virasoro} to this point. 
This, in particular, means that the sources $j_{\rho}(\tau)$ and $j_{f_i}(\tau)$ are not arbitrary functions of $\tau$ but constrained by the Virasoro constraints.

From the viewpoint of the boundary action as the generating functional of correlation functions, the boundary graviton $B(\tau)$ is ``the new kid on the block.'' We would like to clarify what role exactly it plays in the holographic framework.  
As may have been anticipated, the answer is simple and we shall show that the $B(\tau)$ equation of motion of the 1$d$ boundary theory is equivalent to the 2$d$ bulk Virasoro constraints, at least, on the $SL(2,R)$ invariant vacuum:
\begin{align}\label{VSchcorrespondence}
\boxed{{\delta S^{({\rm 1d})}_{\rm eff}[B, j]\over\delta B}=0\qquad\Longleftrightarrow\qquad \left. T^{({\rm 2d})}_{\pm\pm}\right|_{z\to 0}=0}
\end{align}
Since the Virasoro constraints are the equations of motion for $g_{\pm\pm}$ and the boundary graviton $B(\tau)$ is a remnant of 2$d$ metric degrees of freedom, it is not a surprise that this correspondence holds.

We first present the $B(\tau)$ equation of motion of the 1$d$ boundary effective action \eqref{effectiveaction}. The computational details are shown in Appendix \ref{computationsII}. There are three parts in the pseudo Nambu-Goldstone boson action \eqref{pNGfinal} and the variations of each part are given by 
\begin{align}
{\delta S_{j_{\rho}{\rm Sch}}\over \delta B}&=- B'\del^3_B\tilde{j}_{\rho}\ ,\qquad
{\delta S_{j_{\rho}^2}\over \delta B}=-{2\pi i\over 3}B' \left(\tilde{j}_{\rho}\del^4_B\tilde{j}_{\rho}+2\del_B\tilde{j}_{\rho}\del^3_B\tilde{j}_{\rho}\right)\ ,\\
{\delta S_{j_{\rho}^2{\rm Sch}}\over\delta B}&=-2\pi i B' \Biggl[4\del_B\tilde{j}_{\rho}\del^3_B\tilde{j}_{\rho}+3\left(\del^2_B\tilde{j}_{\rho}\right)^2
+\tilde{j}_{\rho} \del^4_B \tilde{j}_{\rho}\nn\\
&\hspace{2.5cm}+\del_B^2\left(\tilde{j}_{\rho}^2{\{B(\tau), \tau\}\over B'^2}\right)-2\tilde{j}_{\rho}\del^2_B \tilde{j}_{\rho}{\{B(\tau), \tau\}\over B'^2}\Biggr]\ ,
\end{align}
where $\tilde{j}_{\rho}(B(\tau))=j_{\rho}(\tau)B'(\tau)$ as appeared before. In the meantime, the variation of the matter action reads
\begin{align}
{\delta S_{j_f^2}\over \delta B}= 2\pi i B'\left(\del_Bj_f\right)^2\ .
\end{align}
These then yield the equation of motion
\begin{align}\label{EOMfinal}
0={\delta S^{({\rm 1d})}_{\rm eff}[B, j]\over\delta B}&=B'\Biggl[{N\over 24}\left\{-2\del^3_B\tilde{j}_{\rho}+2\pi i\left(2\del_B\tilde{j}_{\rho}\del^3_B\tilde{j}_{\rho}+3\left(\del^2_B\tilde{j}_{\rho}\right)^2\right)\right\}-2\pi i \sum_{i=1}^N\left(\del_Bj_{f_i}\right)^2\nn\\
&\hspace{2cm}+2\pi i {N\over 24}\left\{\del_B^2\left(\tilde{j}_{\rho}^2{\{B(\tau), \tau\}\over B'^2}\right)-2\tilde{j}_{\rho}\del^2_B \tilde{j}_{\rho}{\{B(\tau), \tau\}\over B'^2}\right\}\Biggr]\ .
\end{align}
This is the LHS of  \eqref{VSchcorrespondence} and to be compared with the Virasoro constraints \eqref{Virasoro}.
Note that to the linear order the equation of motion is $\del_B^3\tilde{j}_{\rho}=0$ whose solution is
\begin{align}
j_{\rho}(\tau)={\alpha +\beta B(\tau)+\gamma B(\tau)^2\over B'(\tau)}
\end{align}
with constants $\alpha$, $\beta$ and $\gamma$ in agreement with the dilaton $\phi_r$ in the JT model \cite{Maldacena:2016upp} and the non-normalizable mode in  Liouville theory \cite{Mandal:2017thl}.

We now turn to the Virasoro constraints \eqref{Virasoro}. We are only concerned with the fluctuation part of the Virasoro constraints  with a constant dilaton.
As shown in Appendix \ref{computationsII}, the second-order Virasoro constraints at the boundary $z\to 0$ take the form
\begin{align}\label{Virasorofinal}
\hspace{-.2cm}
0=T_{\pm\pm}= -\pi i B'^2\left[{N\over 24}\biggl\{-2\del^3_B\tilde{j}_{\rho}+2\pi i \biggl(2\del_B^3\tilde{j}_{\rho}\del_B\tilde{j}_{\rho}+3(\del_B^2\tilde{j}_{\rho})^2\biggr)\biggr\}-2\pi i \sum_{i=1}^N\left(\del_Bj_{f_i}\right)^2\right].
\end{align}
This is the RHS of  \eqref{VSchcorrespondence}.
Since we turned on the non-normalizable modes in the left-right symmetric way, the left and right energy-momentum tensors are identical at the boundary.

We are now in a position to compare the $B(\tau)$ equation of motion \eqref{EOMfinal} and the Virasoro constraints \eqref{Virasorofinal}.
The two are identical except for the second line of  \eqref{EOMfinal} which are the terms involving the Schwarzian derivatives. 
Since the Schwarzian derivative $\{B(\tau), \tau\}=0$ on the $SL(2,R)$ invariant vacuum, we see that as advertized, the Virasoro/Schwarzian correspondence \eqref{VSchcorrespondence} holds on this vacuum for which $B(\tau)=\tau$ modulo M\"obius transformations \eqref{eq:sl2stran}. This is the most conservative interpretation we offer.

However, we would like to discuss a little more speculative interpretation. It was our expectation and is our sentiment that ultimately, the Schwarzian-dependent terms in the second line of \eqref{EOMfinal} would disappear and the Virasoro/Schwarzian correspondence \eqref{VSchcorrespondence} works on all vacua or for all boundary conditions, i.e. for a generic $B(\tau)$. If these terms were a discrepancy to be resolved, we suspect that they are related to $t_{\pm}$ in \eqref{Virasoro}. As remarked in Footnote \ref{foot:auxiliary}, they can be expressed as $t_{\pm}=\del^2_{\pm}\varphi_{\pm}-(\del_{\pm}\varphi_{\pm})^2$ in terms of the auxiliary field $\varphi$. They have the nonvanishing background values $t_{\pm}=\half\{ B(x^{\pm}), x^{\pm}\}$ with $\varphi_{\pm}=\half\ln B'(x^{\pm})$ which vanish on the $SL(2,R)$ invariant vacuum. 
In our analysis we have been agnostic about potential effects of the auxiliary field $\varphi_{\pm}$ on the boundary action. However, it might be that there is a missed effect and when it is properly taken into account, it cancels the Schwarzian-dependent terms in \eqref{EOMfinal}. 

\section{Discussion}\label{sec:discussion}

From the viewpoint of holography, it is rather remarkable to see that there is a straightforward connection between the bulk Einstein equations (for $g_{\pm\pm}$) and  the boundary equation of motion, which we dubbed the Virasoro/Schwarzian correspondence.
The key to this correspondence is the presence of the dynamical boundary graviton $B(t)$. In the standard holography, the boundary graviton does not make a regular appearance except for the AdS$_3$ case \cite{Brown:1986nw} and the AdS/CFT realization of Randall-Sundrum II \cite{Randall:1999vf} as suggested by Gubser \cite{Gubser:1999vj}. Even in these examples, to our knowledge, the direct bulk-boundary connection of the type \eqref{VSchcorrespondence} has not been realized or formulated. A potential generalization to the AdS$_3$ case can be explored by studying the corresponding $2d$ effective action analogous to the $1d$ Schwarzian action \cite{Cotler:2018zff}. It is, however, worth mentioning that there are attempts to derive the bulk Einstein equations from other perspectives such as the entanglement of boundary CFTs \cite{Jafferis:2015del,Lewkowycz:2018sgn, Lashkari:2013koa,Faulkner:2017tkh}.

That said, as remarked in Section  \ref{sec:correspondence}, we could only show that the Virasoro/Schwarzian correspondence is so far exact on the SL$(2,R)$-invariant vacuum. This may not be entirely satisfactory. However, as we discussed, it could be that the mismatched Schwarzian terms in the second line of \eqref{EOMfinal} disappear upon the inclusion of a subtle effect from the background auxiliary field $\varphi_\pm$ and the Virasoro/Schwarzian correspondence holds true on all vacua. We hope to reach a clear understanding of this point in the near future. 
A somewhat related note is that the two point function of the Schwarzian derivative obtained from the action \eqref{pNGfinal} is structurally almost in the form of the OPE of the 2$d$ energy-momentum tensor, $T(z)T(w) \sim c/2/(z-w)^4 +2T(w)/(z-w)^2 +\partial T(w)/(z-w)$, except that the last term is missing. The absence of this last term might be related to the mismatched Schwarzian terms.

In this paper, we have focused on the gravity side of nearly AdS$_2$ holography. Needless to say, it is very important to gain some understanding of the dual quantum mechanics. An obvious candidate is the SYK model \cite{Kitaev:KITP,Sachdev:1992fk} or its variant. Even though we do not have much to offer on this point, it may be worth commenting on the following observation. The Schwarzian sector of the SYK model with $N$ Majorana fermions takes the form, $S= \frac{N \alpha(q)}{J} \int dt \{B(t),t \}$ with the dimension one coupling $J$ and a constant $\alpha(q)$ which depends on the order $q$ of the interaction. The inverse coupling $1/J$ corresponds to the symmetry breaking scale $j_{\rho}$  \cite{Mandal:2017thl,Maldacena:2016hyu} and one may identify $N$ with the number of massless scalars in the qCGHS model.  
Then the second order actions \eqref{rhotwopoint} and \eqref{rho2Schaction} would correspond to the $1/J^2$ correction to the Schwarzian action. 
However, they do not seem to agree with the $1/J^2$ correction in the SYK model \cite{Jevicki:2016bwu, Jevicki:2016ito, Kitaev:2017awl}, indicating that the dual quantum mechanics may not simply be the SYK model. 

Even though we have not discussed in this paper, the qCGHS model has a larger class of exact solutions with matter. For example, there are exact multi shock wave solutions \cite{hirano2019}. These include an AdS$_2$ counterpart of the shock wave limit of traversable wormholes studied in \cite{Hirano:2019ugo}. In order to describe these shock waves, we need to generalize the non-normalizable modes \eqref{deformab} to the left-right asymmetric sources. In particular, it would be interesting to see if and how the boundary action for a traversable wormhole realizes the GJW construction of traversable wormholes via a double-trace deformation \cite{Gao:2016bin}. In contrast to the prior work \cite{Maldacena:2018lmt}, this would be an example of non-eternal traversable wormholes.

Finally, it is important to understand if and how the qCGHS model can be embedded in higher dimensional black holes. As mentioned earlier, the classical CGHS model arises as the effective two-dimensional theory of extremal dilatonic black holes in four and higher dimensions \cite{Gibbons:1982ih, Gibbons:1987ps, Garfinkle:1990qj, Horowitz:1991cd, Giddings:1991mi, Giddings:1992kn}. It is not immediately clear whether the two-dimensional conformal anomaly has an interpretation in the higher dimensional parent theory.
The current technology of black hole microstate counting is limited to supersymmetric extremal black holes. (See \cite{Zaffaroni:2019dhb} for a recent review.)
It is an open question to account for non-extremal black hole entropy from dual field theory. 
If the qCGHS model can be embedded in higher dimensional black holes, 
one can hope to gain a better understanding of non-extremal black holes along the line of recent developments \cite{Moitra:2018jqs,Nayak:2018qej}. 

\section*{Acknowledgements}

We would like to thank Sam van Leuven for the collaboration in the early stage of this work. We would also like to thank Felix Haehl for many useful discussions.
This work was supported in part by the National Research Foundation of South Africa and DST-NRF Centre of Excellence in Mathematical and Statistical Sciences (CoE-MaSS). YL thanks the support from the Department of Science and Technology and the National Research Foundation's South African Research Chairs Initiative for post-doctoral support and the project “Towards a deeper understanding of black holes with non-relativistic holography” of the Independent Research Fund Denmark (grant number DFF-6108-00340). 
Opinions expressed and conclusions arrived at are those of the authors and are not necessarily to be attributed to the NRF or the CoE-MaSS.

\appendix
\section{Non-normalizable modes near boundary}\label{computations}

In this appendix we show the details of the computation for the results in Sections \ref{sec:secondorder} and \ref{Sec:Action}.
There are a few subtleties in the computation and we will clarify them along the way.

\paragraph{Liouville fluctuations:}
In the Poincar\'e coordinates, the first order Liouville fluctuation \eqref{rho1general} takes the form
\begin{align}
\rho_1=-\int_{-\infty}^{\infty} d\tau_0{2z^2 j_{\rho}(\tau_0)\over \left((\tau-\tau_0)+iz\right)^2\left((\tau-\tau_0)-iz\right)^2}
=-\int_{-\infty}^{\infty} du{2z^2 j_{\rho}(\tau+u)\over \left(u+iz\right)^2\left(u-iz\right)^2}\ ,
\end{align}
where we performed a Wick-rotation, $t\to i\tau$ and $j_{\rho}(t)\to -i\,j_{\rho}(\tau)$. 
We now use a contour integral to evaluate the $u$-integral. In order to ensure the convergence of the integral, we adopt the prescription to add a damping factor $e^{i\epsilon (\tau+u)}$ to the source $j_{\rho}(\tau+u)$ and send $\epsilon\to 0_+$ in the very end of the calculation. This selects the preferred contour $C=R\cup H_+$ to be the real axis $R$ plus a semi-circle $H_+$ going around in the UHP.
With this prescription assumed, we find that 
\begin{align}
\rho_1&=-\pi \left[{1\over z}j_{\rho}(\tau+iz)-ij_{\rho}'(\tau+iz)\right]=-\pi\left[{j_{\rho}(\tau)\over z}+\half z j_{\rho}''(\tau)\right]+O(z^2)\ .
\end{align}
It then follows that 
\begin{align}
\rho_1^2={\pi j_{\rho}(\tau)\over z}\int_{-\infty}^{\infty} d\tau_0{2z^2 j_{\rho}(\tau_0)\over \left((\tau-\tau_0)^2+z^2\right)^2}+\mathcal{O}(z^2)
=2\pi z\int_{-\infty}^{\infty}  d\tau_0{j_{\rho}(\tau)j_{\rho}(\tau_0)\over (\tau-\tau_0)^4}+\mathcal{O}(z^2)\ .
\end{align}
Turning to the next order, the second order Liouville fluctuation \eqref{rho2general} consists of three terms
\begin{align}
\rho_2=\rho^{\rm I}_{2+}+\rho^{\rm I}_{2-}+\rho^{\rm II}
\end{align}
where in the Poincar\'e coordinates
\begin{align}
\rho^I_{2\pm}=-{1\over 4}\left(\int_{-\infty}^{\infty}d\tau_0{j_{\rho}(\tau_0)\over \left((\tau-\tau_0)\pm iz\right)^2}\right)^2\quad\mbox{and}\quad
\rho^{\rm II}_{2}&=\half\left(\int_{-\infty}^{\infty}d\tau_0{j_{\rho}(\tau_0)\over (\tau-\tau_0)^2+z^2}\right)^2\ .
\end{align}
With the above prescription of the damping factor for the source $j_{\rho}(\tau_0)$, we perform one $\tau_0$-integral for each term by using the contour integral along $C$ as was done for the first order fluctuation $\rho_1$. These integrals result in
\begin{align}
\rho^I_{2+}=-{\pi i\over 2} j'_{\rho}(\tau+ iz)\int_{-\infty}^{\infty}d\tau_0{j_{\rho}(\tau_0)\over \left((\tau-\tau_0)+ iz\right)^2}\ ,\qquad\qquad
\rho^I_{2-}=0\ ,
\end{align}
and
\begin{align}
\rho^{\rm II}_{2}={\pi \over 2z}j_{\rho}(\tau+iz)\int_{-\infty}^{\infty}d\tau_0{j_{\rho}(\tau_0)\over (\tau-\tau_0)^2+z^2}\ .
\end{align}
For a small $z$ we then find that
\begin{align}
\rho_2= {\pi\over 2}\int_{-\infty}^{\infty}  d\tau_0\left[{1\over z}{j_{\rho}(\tau)j_{\rho}(\tau_0)\over (\tau-\tau_0)^2}
+z\left({\del^2\over\del\tau^2}\left(\half{j_{\rho}(\tau)j_{\rho}(\tau_0)\over (\tau-\tau_0)^2}\right)-{4j_{\rho}(\tau)j_{\rho}(\tau_0)\over (\tau-\tau_0)^4}\right)\right]
+\mathcal{O}(z^2).
\end{align} 

\paragraph{Massless matter:} 
In the Poincar\'e coordinates, the massless matter non-normalizable mode \eqref{NNmatter} takes the form 
\begin{align}
f_i=\int_{-\infty}^{\infty}d\tau_0{2z j_{f_i}(\tau_0)\over (\tau-\tau_0)^2+z^2}\ ,
\end{align}
where we performed a Wick-rotation $t\to i\tau$ and $j_{f_i}(t)\to -i\,j_{f_i}(\tau)$. As in the case of the Liouville fluctuations, we adopt the prescription to add a damping factor $e^{i\epsilon\tau_0}$ to the sources $j_{f_i}(\tau_0)$ and use the contour integral along $C$ to calculate $f_i$. We then find 
\begin{align}
f_i=2\pi  j_{f_i}(\tau+iz)\ .
\end{align}
It then follows that
\begin{align}
f_i^2=4\pi z\int_{-\infty}^{\infty}d\tau_0{ j_{f_i}(\tau)j_{f_i}(\tau_0)\over (\tau-\tau_0)^2}+\mathcal{O}(z)\ .
\end{align}
Taking the derivative with respect to $z$, the covariantization of the expression yields the matter action in \eqref{matteraction}.

\section{Variations of boundary action and Virasoro constraints}\label{computationsII}

Here we show the computational details of the $B(\tau)$ equation of motion of the boundary theory and the Virasoro constraints to the second order in the Liouville fluctuation as discussed in Section \ref{sec:correspondence}.

\paragraph{The $B(\tau)$ equation of motion:}
The variation of the Schwarzian action \eqref{Schwarzianaction} with respect to $B(\tau)$ is given by
\begin{align}\label{LiouvilleEOM1st}
\delta_BS_{j_{\rho}{\rm Sch}}&=\delta_B\int d\tau j_{\rho}(\tau)\,\frac{B'''(\tau) B'(\tau)-{3\over 2} B''(\tau)^2}{ B'(\tau)^2}
=-\int dB(\tau)\delta B(\tau){\del^3\left(j_{\rho}(\tau)B'(\tau)\right)\over \del B(\tau)^3}\nn\\
&=-\int dB(\tau)\delta B(\tau)\del^3_B\tilde{j}_{\rho}\ .
\end{align}
The variation of the first quadratic part \eqref{rhotwopoint} is calculated as
\begin{align}
\delta_BS_{j_{\rho}^2}&=\delta_B\int d\tau\int d\tau_0{B'(\tau)^{2}B'(\tau_0)^{2}j_{\rho}(\tau)j_{\rho}(\tau_0)\over (B(\tau)-B(\tau_0))^{4}}\nn\\
&=8\int dB(\tau)\delta B(\tau) \left(j_{\rho}(\tau) B'(\tau)\right)\int dB(\tau_0) \frac{j_{\rho}(\tau_0)B'(\tau_0) }{(B(\tau)-B(\tau_0))^5}\nn\\
&\hspace{1cm}-4\int dB(\tau)\delta B(\tau){\del\left( j_{\rho}(\tau) B'(\tau)\right)\over\del B(\tau)}\int dB(\tau_0)\frac{j_{\rho}(\tau_0)B'(\tau_0) }{(B(\tau)-B(\tau_0))^4}\nn\\
&=\int dB(\tau)\delta B(\tau)\left[ {2\pi i\over 3}\left(-\tilde{j}_{\rho}\del^4_B\tilde{j}_{\rho}-2\del_B\tilde{j}_{\rho}\del^3_B\tilde{j}_{\rho}\right)\right]\ ,
\label{2ndrhoEOM}
\end{align}
where we used integration by parts and adopted the prescription for the $B(\tau_0)$-integral
\begin{align}
\int_{-\infty}^{\infty} dB(\tau_0)\frac{\tilde{j}_{\rho}(B(\tau_0))}{(B(\tau)-B(\tau_0))^n}=(-1)^n{2\pi i\over (n-1)!} \del_B^{n-1}\tilde{j}_{\rho}\left(B(\tau)\right)\ .
\label{Btau0integral}
\end{align}
For the variation of the second quadratic part \eqref{rho2Schaction}, the computation is tedious but it can be calculated as
\begin{align}
&\delta_BS_{j_{\rho}^2{\rm Sch}}
=\int d\tau \int d\tau_0{j_{\rho}(\tau)j_{\rho}(\tau_0)B'(\tau_0)\over (B(\tau)-B(t_0))^3 B'(\tau)^3}\nn\\
&\times\Biggl[B'(\tau) \left(3 B''(\tau)^2-2 B'''(\tau) B'(\tau)\right) \biggl\{B'(\tau_0) (\delta B(\tau)-\delta B(\tau_0))
+(B(\tau_0)-B(\tau))\delta B'(\tau_0)\biggr\}\nn\\
&+(B(\tau)-B(\tau_0)) B'(\tau_0) \biggl\{\delta B'''(\tau) B'(\tau)^2-3 \delta B''(\tau)B'(\tau) B''(\tau) 
+\delta B'(\tau) \left(3 B''(\tau)^2-B'''(\tau) B'(\tau)\right)\biggr\}\Biggr]\ .\nn
\end{align}
Performing integration by parts and using $\tilde{j}_{\rho}(B(\tau))=j_{\rho}(\tau)B'(\tau)$, this can be rewritten as
\begin{align}
&\delta_BS_{j_{\rho}^2{\rm Sch}}
=\int dB(\tau) \delta B(\tau)\int dB(\tau_0)  \tilde{j}_{\rho}(\tau_0) \Biggl[-\frac{\del_B\tilde{j}''_{\rho}(\tau)}{(B(t)-B(\tau_0))^2 B'(t)^2}\nn\\
&+\del_B\tilde{j}'_{\rho}(\tau) \left\{\frac{6}{(B(\tau)-B(\tau_0))^3 B'(\tau)}+\frac{3B''(\tau)}{(B(\tau)-B(\tau_0))^2 B'(\tau)^3}\right\}
-\del_B\tilde{j}_{\rho}(\tau) \biggl\{\frac{18}{(B(\tau)-B(\tau_0))^4}\nn\\
&+\frac{6B''(\tau)}{(B(\tau)-B(\tau_0))^3 B'(\tau)^2}+\frac{3B''(\tau)^2-B'''(\tau)B'(\tau)}{(B(\tau)-B(\tau_0))^2 B'(\tau)^4}
-\frac{3 B''(\tau_0)^2-2 B'''(\tau_0) B'(\tau_0)}{(B(\tau)-B(\tau_0))^2B'(\tau_0)^4}\biggr\}\nn\\
&+\tilde{j}_{\rho}(\tau) \left\{\frac{24}{(B(\tau)-B(\tau_0))^5}+\frac{3 B''(\tau)^2-2 B'''(\tau) B'(\tau)}{(B(\tau)-B(\tau_0))^3 B'(\tau)^4}+\frac{3 B''(\tau_0)^2-2 B'''(\tau_0) B'(\tau_0)}{(B(\tau_0)-B(\tau))^3B'(\tau_0)^4}\right\}
\Biggr]\ .\nn
\end{align}
Adopting the prescription for the $B(\tau_0)$-integral \eqref{Btau0integral}, after a little manipulations, we finally obtain that 
\begin{align}
\delta_BS_{j_{\rho}^2{\rm Sch}}=2\pi i\int dB(\tau) \delta B(\tau) \Biggl[&-4\del_B\tilde{j}_{\rho}\del^3_B\tilde{j}_{\rho}-3\left(\del^2_B\tilde{j}_{\rho}\right)^2
-\tilde{j}_{\rho} \del^4_B \tilde{j}_{\rho}\nn\\
&-\del_B^2\left(\tilde{j}_{\rho}^2{\{B(\tau), \tau\}\over B'^2}\right)+2\tilde{j}_{\rho}\del^2_B \tilde{j}_{\rho}{\{B(\tau), \tau\}\over B'^2}\Biggr]\ .
\end{align}
Finally, using again the prescription \eqref{Btau0integral}, the variation of the matter action reads
\begin{align}\label{matterEOM}
\delta_BS_{j_f^2}&=-\int d\tau \delta B(\tau)  j_f'(\tau)\int d\tau_0 \frac{ j_f(\tau_0) B'(\tau_0)}{(B(\tau)-B(\tau_0))^2}
= -2\pi i \int dB(\tau) \delta B(\tau) \left(\del_Bj_f\right)^2\ .
\end{align}

\paragraph{Virasoro constraints:}
The linear fluctuation part of the Liouville energy-momentum tensor is found to be
\begin{align}
T^{(1)}_{\pm\pm}&\equiv -{N\over 12}\left[2\del_{\pm}\rho_0\del_{\pm}\rho_1-\del^2_{\pm}\rho_1\right]
={N\over 4}B'(x^{\pm})^2\int_{-\infty}^{\infty}dB(\tau_0)\frac{ \tilde{j}_{\rho}(B(\tau_0))}{(B(x^{\pm})-B(\tau_0))^4}\nn\\
&\stackrel{z\to 0}{\longrightarrow} {2\pi iN\over 24}B'^2\del^3_B\tilde{j}_{\rho}\ .
\end{align}
The second-order fluctuation part is calculated as
\begin{align}
T^{(2)}_{\pm\pm}&\equiv -{N\over 12}\left((\del_{\pm}\rho_1)^2+2\del_{\pm}\rho_0\del_{\pm}\rho_2-\del^2_{\pm}\rho_2\right)\nn\\
&=-{N\over 4}B'(x^{\pm})^2\Biggl[\int_{-\infty}^{\infty}{dB(\tau_0)\tilde{j}_{\rho}(B(\tau_0))\over \left(B(x^{\pm})-B(\tau_0)\right)^4}
\int_{-\infty}^{\infty}{dB(\tau_0)\tilde{j}_{\rho}(B(\tau_0))\over \left(B(x^{\pm})-B(\tau_0)\right)^2}
+\left(\int_{-\infty}^{\infty}{dB(\tau_0)\tilde{j}_{\rho}(B(\tau_0))\over \left(B(x^{\pm})-B(\tau_0)\right)^3}\right)^2\Biggr]\nn\\
&\stackrel{z\to 0}{\longrightarrow}-{N\over 48}(2\pi i)^2B'^2\left[2\del_B^3\tilde{j}_{\rho}\del_B\tilde{j}_{\rho}+3(\del_B^2\tilde{j}_{\rho})^2\right]\ .
\end{align}
Finally, the matter energy-momentum tensor is found as
\begin{align}\label{matterstress}
T^f_{\pm\pm}&=\half\sum_{i=1}^NB'(x^{\pm})^2\int_{-\infty}^{\infty} {dB(\tau_0)\tilde{j}_{f_i}(B(\tau_0)))\over (B(x^{\pm})-B(\tau_0))^2}
\int_{-\infty}^{\infty} {dB(\tau'_0)\tilde{j}_{f_i}(B(\tau'_0)))\over (B(x^{\pm})-B(\tau'_0))^2}\nn\\
&\stackrel{z\to 0}{\longrightarrow}{(2\pi i)^2\over 2}B'^2\sum_{i=1}^N\left(\del_Bj_{f_i}\right)^2\ .
\end{align}

\bibliographystyle{JHEP}
\bibliography{solutions}

\end{document}